\documentclass[11pt,preprint]{aastex}
\usepackage[english]{babel}

\RequirePackage{ifmtarg}
\def\hdr@do@rule[#1]#2#3{\rule[#1]{#2}{#3}\let\hdr@do@something=\hdr@do@skip}
\def\hdr@do@skip[#1]#2#3{\hspace*{#2}\let\hdr@do@something=\hdr@do@rule}
\usepackage{dashrule}
\usepackage{lipsum}
\usepackage[framemethod=TikZ]{mdframed}
\mdfdefinestyle{mystyle}{%
frametitlerulewidth=0pt,
frametitlerule=true,
framelinestyle=dashed,
frametitlefont=\color{black},
frametitlebackgroundcolor=yellow,
}

\usepackage[colorlinks=true,linkcolor=blue,citecolor=blue]{hyperref}%

\usepackage[utf8]{inputenc}
\usepackage[english]{babel}
\usepackage{imakeidx}
\usepackage{longtable}
\usepackage{booktabs}
\usepackage{threeparttablex}
\usepackage{color}
\usepackage{fancyhdr} 
\usepackage{tikz}
\usepackage{showexpl}
\usepackage{wrapfig}

\makeindex

\usepackage[framemethod=TikZ]{mdframed}
\usepackage{lipsum}

\mdfdefinestyle{MyFrame}{%
    linecolor=magenta,
    outerlinewidth=3pt,
    roundcorner=10pt,
    innertopmargin=\baselineskip,
    innerbottommargin=\baselineskip,
    innerrightmargin=20pt,
    innerleftmargin= 20pt,
    backgroundcolor=white, 
    linecolor=green,
    linewidth=2pt,
    frametitlerule=true,
    outerlinewidth=1pt,
    middlelinewidth=2pt,
    innerlinecolor=green,
    middlelinecolor=yellow,
    innerlinewidth=1pt}

    \mdfdefinestyle{pracMyFrame}{%
    linecolor=red,
    innerlinecolor=red,
    middlelinecolor=yellow,
    roundcorner=-18pt,
    innertopmargin=\baselineskip,
    innerbottommargin=\baselineskip,
    innerrightmargin=20pt,
    innerleftmargin= 20pt,
    backgroundcolor=cyan!20,
    linewidth=2pt,
    frametitlerule=true,
    outerlinewidth=2pt,
    middlelinewidth=2pt,
    innerlinewidth=2pt
   }
 
\newif\ifcolor
\colorfalse

\mdfsetup{skipabove=\topskip,skipbelow=\topskip}
\newcounter{sidebarnumber}
\newcounter{pracsidebarnumber}

\global\mdfdefinestyle{exampledefault}{pstrickssetting={dash pattern = on 10pt off 3pt,},linecolor=blue,linewidth=5pt}

\hypersetup{
    colorlinks=true,
    linkcolor=blue,
    filecolor=magenta,      
    urlcolor=cyan,
}

\newif\ifisapj
\isapjfalse

\begin{document}

\thispagestyle{empty}
\setcounter{sidebarnumber}{0}
\setcounter{pracsidebarnumber}{0}

\thispagestyle{empty}
\setcounter{sidebarnumber}{0}
\setcounter{pracsidebarnumber}{0}

\begin{center}
{\Large 
Studies in Astronomical Time Series Analysis: VII.\\
An Enquiry Concerning  
Non-Linearity, the RMS-Mean Flux Relation,\\
and log-Normal Flux Distributions
}

\vskip 0.5in
\Large
Jeffrey D. Scargle\\
Astrobiology and Space Science Division\\
NASA Ames Research Center\\
Jeffrey.D.Scargle@nasa.gov

\end{center}

\ifisapj \else

\begin{quotation}
``... when we have often employed any term,
though without a distinct meaning, 
we are apt to imagine it has a determinate idea
annexed to it.  ... 
When we entertain, therefore, any suspicion,
that a philosophical term is employed without any meaning
or idea (as is but too frequent), we need but
enquire, {\it from what impression is that supposed idea derived?''}\\
\\
David Hume, An Enquiry Concerning Human Understanding,
1748 
\end{quotation}

 \textbf{\emph{
\begin{quotation}
``If language is not correct, then what is said is not what is meant; if what is said is not what is meant, then what must be done remains undone; if this remains undone, morals and art will deteriorate; if justice goes astray, the people will stand about in helpless confusion. Hence there must be no arbitrariness in what
is said.  This matters
above everything.''\\
\\
Confucius, on the ``rectification of names,''
Analects, Book 13, Verse 3, ca. 500 B.C. 
(Translation by James R. Ware)
\end{quotation}}}

\fi

\begin{abstract}

A broad and widely used 
class of stationary,
linear, additive time series models
can have statistical properties 
which many authors have asserted 
imply that the underlying process 
must be non-linear, non-stationary,
multiplicative, or 
inconsistent with shot noise.
This result is demonstrated 
with exact and numerical evaluation of
the model flux distribution function 
and dependence of flux 
standard deviation on mean flux
(here and in the literature called
the \emph{rms-flux relation}).
These models can:
(1) exhibit  normal, log-normal or 
other flux distributions;
(2) show linear or 
slightly non-linear 
rms-mean flux dependencies;
as well as (3) 
match arbitrary second order
statistics of the time series data.
Accordingly the above assertions 
cannot be made on the basis of 
statistical time series analysis alone.
Also discussed are 
ambiguities in 
the meaning of terms relevant 
to this study -- 
\emph{linear}, 
\emph{stationary} and 
\emph{multiplicative} -- 
and functions that can 
transform observed fluxes 
to a normal distribution 
as well or better than the logarithm.
\end{abstract}

\ifisapj \else
\tableofcontents
\clearpage \fi

\section{Introduction}
\label{introduction}

A widespread goal in the study of 
stochastic variation of astronomical systems 
is to explore underlying physical processes,
which are unfortunately not directly observable.
In a seminal paper 
\cite{press} coined the evocative
term \emph{flickering} 
(now often 
called \emph{$1/f$ noise} or \emph{colored noise}), and 
commented that 
explanation of 
physical and astrophysical processes 
variable on all timescales 
had been exceedingly difficult.
There has since been considerable progress,
for example 
through improved data analysis methods 
\cite[e.g.][]{vaughan}
and through study of explicitly non-linear, 
chaotic dynamical models 
\citep[e.g.][]{scargle_dhr,mineshige,aschwanden}.

However, the inner workings of 
astrophysical systems 
remain largely mysterious.
This fact is
probably not due to 
a data shortage,
in view of the great advances 
in observational capabilities
in modern astronomy.
Elucidating complex, dynamic, three-dimensional 
astrophysical systems -- 
with uncertain physical processes and parameters --
from one-dimensional time series data
is intrinsically difficult.
But an important added factor is
the tortuous path from time series data 
to inferred dynamical properties, 
strewn with misunderstandings
about the nature of their connections.

It is the purpose of this note to 
clarify some of these problematic issues
in analyzing distributions of
observed fluxes,
and dependencies between
flux and flux variance.
A new derivation of properties of 
a broad class of linear, stationary, and additive 
processes demonstrates that 
they can possess 
linear and near-linear variance-mean flux 
relations and arbitrary flux distributions,
including the range from normal to log-normal and beyond.
Therefore such statistical properties of light curves 
should not be taken to indicate the presence of 
non-linearity, non-stationarity, or multiplicativity, 
or to disallow the presence of shot-noise characteristics,
as frequently asserted.

A partial summary of previous work is 
followed by 
clarification of how the relevant 
terms \emph{linear}, 
\emph{stationary}, 
\emph{additive}, 
and their opposites, 
are  used here.
Subsequent sections 
briefly define standard 
autoregressive/moving average 
(ARMA)
random process models, and
then 
analyze the rms vs. mean flux
relation 
and flux-distribution properties 
of the models.
Extensions  to incorporate 
some forms of nonstationarity,
short- and long- term memory,
and fractional Brownian motion,
beyond the scope here,
are worth further study --
e.g. the 
autoregressive fractionally integrated moving average
(ARFIMA) models discussed in
\cite{granger_joyeux}.


\section{Previous Work}
\label{previous}

\cite{denis} were apparently 
the first to report ``source noise flux''
linearly increasing with ``source total flux,'' 
using x-ray observations of  Nova Persei.
The influential paper
\cite{uttley_mchardy} 
elaborated this idea,
importantly broadening the class 
of sources that demonstrate a 
linearly increasing 
dependence of root-mean-square 
variability on mean flux.
Detailed 
X-ray light curves of Cyg X-1 and SAX J1808.4-3658 
showed a remarkably tight linear relation,
and three Seyfert galaxies showed similar
dependence, albeit with crude flux resolution.

Since this initial work,  
the linear ``rms vs flux'' relation 
has been extended by various authors 
to many sources and source classes,
leading to the attribution of near 
ubiquity to this relation.
A key paper \citep{uttley_mchardy_vaughan}
reported a detailed analysis in the context of
X-ray binaries and active galaxies.
\cite{vaughan_uttley}
discuss tests for non-linearity, non-Gaussianity, 
and time asymmetry 
using statistics beyond second order.
\cite{giebels_degrange}
commented on the approximately log-normal 
flux distribution and a relatively 
scattered rms-mean flux relation in BL Lacertae.
\cite{heil_vaughan_uttley}
found the rms-flux relation 
present in several black hole binaries,
with systematic dependence of the
slope and intercept on hardness state.
\cite{scaringi_1} found 
linear rms-flux relations in 
Kepler data for the white dwarf MV Lyrae.
\cite{dobrotka_ness}
looked for a rms-flux relation in Kepler 
data for V1504 Cyg,
finding it in quiescent 
time intervals and, in modified form, 
in outbursts.
\cite{kushwaha} and 
\cite{shah} report log-normal flux distributions
in Fermi Gamma-Ray Space Telescope (FGST) blazar data.
\cite{alston} 
studied non-stationarity and other 
time series properties using simulations.

Some work has attempted to 
link relevant observations with 
theoretical models.
The study of accretion-disk fluctuations by
\cite{lyubarskii},
while not directly addressing  the 
issues discussed here,
has influenced some work that does.
\cite{hogg_reynolds} 
studied propagating fluctuations 
in MHD models of turbulent disk accretion, 
in connection with log-normalcy 
and linear rms-mean flux relations.
\cite{phillipson} compared topological 
features of return maps of 
X-ray light curves of the  binary 4U1705-44 
and those of a system exhibiting non-linear
chaotic behavior.
\cite{sinha} addressed the possibility that  
linear Gaussian variations of particle acceleration
and escape times 
can produce non-Gaussian flux distributions, 
including log-normal ones.
\cite{dobrotka} studied a model for fast variability. \cite{bhatta_dhital} find 
log-normal flux distributions
and linear rms-mean flux relations
in Fermi gamma-ray data for 
20 blazars,
discussing 
possible contact with
models with propagating 
relativistic shocks.

Recent work includes 
identification of non-linear rms-mean flux 
(that is, with some curvature),
in sources and wavelengths 
where neither 
rms-mean flux relations 
nor log-normality are present, 
and a simple model for 
linear rms-mean flux 
relations.
\cite{edelson}
displayed 
non-linear rms-flux relations
in Kepler data for 
the BL Lac galaxy W2R1926+42.
\cite{krista}, in a study of 21 AGN 
Kepler optical light curves, 
found neither log-normal 
flux distributions nor rms-flux relations;
Smith adds that there is no evidence 
for an rms-flux relation in any analysis of 
the best-studied Kepler AGN
Zw229-15 in particular 
(private communication).
\cite{alston_et_al} 
find a non-linear
rms-flux dependence 
rms $ \propto $ flux$^{2/3}$
in x-ray time series 
for the Seyfert galaxy
IRAS 13224-3809.
Of course it is hard to know 
to what extent there have been studies
where relevant negative results were not 
published.
\cite{koen} has proposed 
a model in which the rms-mean flux
relation is due to simple scaling effects,
spurring a response by \cite{uttley_4}
asserting that without modification
Koen's model does not 
yield linear rms-flux
relations on a wide range of time scales 
or log-normal flux distributions.

These and other authors have 
advanced a variety of 
often conflicting conjectures 
about underlying physical processes,
based on statistical characteristics 
of light curves.
Below evidence against 
such conjectures is provided 
by the result that the relevant 
attributes are produced by 
simple, general,
and naturally motivated statistical models 
-- without reference to specific physical mechanisms,
nor any element of 
non-linearity nor of 
non-stationarity 
nor of multiplicativity.
In some cases more recent authors 
seem to have misunderstood or ignored 
caveats in the foundational work in, e.g. 
\cite[][Appendix D]{uttley_mchardy_vaughan} 
and \cite{uttley_4}.
Any criticism implied by the discussion
below is aimed at the 
tangled
web of interactions between various 
ideas, and not at specific authors.

\section{Time Series Descriptors }
\label{descriptors}

It is hoped that the reader will 
excuse the didactic nature of this discussion,
in view of confusions in terminology 
describing statistical properties 
of time series permeating the literature.
In view of the importance of 
clarity in scientific communication,
ambiguous 
short-hand terminology -- 
especially in scientific publications, but 
even in informal settings 
such as white-board discussions -- 
are appropriate only if 
all participants understand the 
same meanings.

In regimes of non-linear growth 
of structures, cosmologists call 
the absolute square of the (linear) 
Fourier transform of the 
mass density field
the \emph{non-linear power spectrum}.
Meant as a convenient shorthand,
this usage can be misleading in several ways.
It uses a term defined in one 
domain (gravitational clustering of matter) 
in a qualitatively different domain 
(second-order statistics of spatial data),
i.e. applying a physics-based descriptor 
to something non-physical.
And it suggests that unambiguous  
signatures of nonlinear physics
can appear directly in the power spectrum.

This section
addresses terms 
describing properties 
of mathematical models  
or physical processes.
Importing these concepts 
to time series data analysis,
albeit from these well-defined settings, 
can be fraught with vagueness, ambiguity and 
confusion.
When there is more than one
applicable meaning,  
short-hand terminology is confusingly ambiguous 
without clear definition of 
the sense intended.
Special circumstances 
and assumptions 
necessary in the mathematical 
or physical contexts  
can be forgotten, ignored, 
or insufficiently understood 
as applied to data.
The imported concept may 
refer to statistics that 
cannot be directly estimated from the 
data alone.

The following sub-sections 
elaborate common threads in 
three yin-yang pairs:
linear/non-linear, stationary/non-stationary,
and additive/multiplicative,
in each case attempting to describe
how physical or mathematical 
properties can give one the impression 
that the term applies to 
time series data.
Similar considerations 
apply to other dualities,
such as causal/acausal,
minimum/maximum delay,
time-reversal invariant/non-invariant, and
analytic/non-analytic \citep{javi_rafa_1,javi_rafa_2},
but are not discussed here.

\subsection{Is \emph{Linearity} Meaningful for Time Series?}
\label{linearity}

The concepts of linearity and non-linearity,
well defined in mathematics
and some physics contexts,
have seeped into areas of astrophysics 
where they are not well defined.
In the context of 
random flickering in astronomical light curves, 
linearity is a property of processes
(applying to mathematical
or physical systems) but not to time series data.
To describe light curves 
as linear or non-linear, without
extension or generalized definition, is a category 
error\footnote{The interesting  page  
\emph{plato.stanford.edu/entries/category-mistakes/}
at the
Stanford 
Encyclopedia of Philosophy
gives a  pertinent example: ``The number two is blue.''
Similar logical problems beset 
the application of 
physics-based concepts of nonlinearity 
to time series data.
As in the above Hume quotation, 
the goal here is 
to enquire from what impressions 
the supposed idea of 
nonlinear time series data 
is derived.
}
-- ascribing to something 
a property that by 
definition cannot apply to it.
These facts are recognized in some quarters,
but terms like ``linear data''
and ``non-linear time series'' 
appear frequently in the literature,
often without 
qualification or explanation.
Approaches range
from taking these terms 
to be so unambiguous
that the plain meaning 
rule\footnote{In law,
if the language of a statute or contract
is unambiguous and clear on its face, 
its meaning must be determined from this 
language and not from extrinsic evidence,  
subject to a 
limitation if the rule leads to absurdity. }
applies and 
no definition is required, 
to thoughtful consideration of clear definitions.
Unfortunately the latter 
is much rarer than the former.
Accordingly 
it is useful to consider three contexts, as follows:

{\bf Mathematics: }
Here the concept is simple:
function $F(x)$ is \emph{linear} in its 
argument if 
\begin{eqnarray}
F(  x + y )&=& F( x ) + F( y ) \\
F(  a x )&=& a F( x )
\label{linear_math}
\end{eqnarray}
for arbitrary $x$, $y$ and $a$.
Usages in other contexts -- e.g. 
 \emph{linear term} characterizing 
the first order part of a 
Taylor series expansion, or 
a term in an equation 
proportional to the independent variable --
derive from this relation.

{\bf Physics: }
This mathematical concept 
can be extended to only those 
physical systems with 
a clearly identified and quantitative 
input/output property -- corresponding
to $F$ mapping an input into 
an output.
A textbook example is a spring:
an applied force (input)
produces a stretch of the spring (output).
In Physics 1 we learn Hooke's law:
within limits the displacement
(output $F$) 
is linear in the applied force (input $x$)
but becomes non-linear with larger force 
if there are corrections to 
Equation (\ref{linear_math}).
This concept rarely applies to complicated systems 
as a whole, and is undefined unless it is  
possible to identify an idealized subsystem
with the input-output property.
Even then it applies only to 
that subsystem, which 
may or may not be 
a primary determinant of 
an observable such as
the time evolution of  flux.
In theoretical physics 
one encounters much the same 
mathematical concepts described above.

{\bf Astrophysics: }
Even a complex astronomical object  
typically has a well defined output:
its emitted flux.
If there is a ``mechanism''
turning some ``input'' into 
all or part of this flux, 
both of these are largely unknown
(or hypothetical in the case of 
a physical model).
Indeed, the
goal of the analysis is to 
identify and elucidate these.
Most often there are many 
mechanisms, or sub-processes, 
interacting with each other in various ways.
Some of these are 
perhaps describable as 
input-output systems, others not.
It seems unlikely that unambiguous 
linearity/non-linearity signatures
due to a single ``central engine''  
will appear commonly 
in light curves for such systems.
Nevertheless, 
let us explore some 
possible ways to attribute meaning to 
such properties 
in time series data.

Perhaps the simplest idea springs from 
fitting simple parametric 
time-domain models directly
to light curves,
as simple as linear trends.
Comparison of descriptions of the data 
using linear and non-linear 
regression can obviously be made to yield 
definitions of linearity or non-linearity.
While this concept is only indirectly 
related to dynamics, 
it may be what some analysts 
have in mind.

Another approach is to 
compare descriptions using 
dynamics-motivated 
linear and non-linear models
of the time evolution of the observable.
The entire book by 
\cite{priestley} is devoted to this viewpoint.
Time series data better described 
by models of the latter class could be 
said to be non-linear.
\cite{brockwell_davis} 
frame linearity in terms of 
equivalence of 
optimal and linear prediction.
An example of a linear model is
the autoregressive/moving average 
process described below.
The Volterra process  
\citep{priestley,uttley_mchardy_vaughan} -- essentially a 
Taylor series expansion generalizing 
the autoregressive formalism with 
a potentially infinite series of explicitly nonlinear terms -- 
is an example of a non-linear model.
There is a fundamental problem 
with applying this approach to stationary processes:
The Wold Theorem guarantees the
existence of a linear model exactly 
representing any stationary
process, even if it is
in some sense putatively non-linear.
That is, 
within the context of 
stationarity, 
these models ``fit'' the data precisely as well as 
Volterra models -- or 
any other non-linear 
model for that matter.

Another obvious problem of this enormous
generalization is nicely described by
 \cite{granger_andersen}:
``At first sight, there may seem to be an 
overpowering richness of possibilities 
once the linear constraint on models 
is removed.'' 
These authors slightly ameliorate 
this difficulty by adding 
`` ... but if certain sensible 
restrictions
are placed on the models,
very many of the possibilities can be 
removed.'' 
(One such restriction is 
the admittedly subjective 
constraint that models have intuitive appeal.
Another is not 
``exploding off to infinity at an exponential 
rate." 
A third is a subtle concept called  \emph{invertibility}.) 
Comparison of
linear against non-linear models
will be
dependent on the classes of models 
of each form considered, 
potentially leading 
to much uncertainty of the results.
Theoretical constraints or other 
considerations -- such as \emph{parsimony}, 
an important model simplicity principle --  may 
reduce the size and complexity of the
model space.

Yet another problem is the need for 
a quantitative goodness-of-fit measure 
of some sort to compare models.
Simple mean error measures are 
not necessarily applicable:
since autoregressive models  
reproduce the data exactly 
(see Section \ref{ar_ma_models}),
model quality is assessed via properties of 
the random driving process 
(called the \emph{innovation}; see below).
Comparison of models using different
quantitative measures is obviously
fraught with difficulties.

A somewhat different approach is 
to attempt to measure non-linearity 
in the form of a metric associated with 
non-linear (``chaotic'') dynamics
\citep{tong,theiler,buchler,sprott}.
In addition to the fact that such 
analyses typically rely on very long 
sequences of high signal-to-noise 
data (e.g., to ensure many near-returns
to the same state),
rare in astronomy, 
there are fundamental estimation 
problems as described by 
\cite{osborne_provenzale,
ruelle,eckmann_ruelle}.
\cite{theiler}
discuss other practical 
difficulties and caveats with the 
use of surrogate data in this context.
Some recent developments \cite[][e.g.]{phillipson}
in this area show progress at overcoming 
such limitations.

An approach, 
often implicit rather than clearly defined,
is to interpret large amplitude flares 
as signatures of non-linearity.
Of course 
``small amplitude = linear; large amplitude = non-linear'' 
can, with care, 
be turned into an unambiguous criterion.
However its connection to the mathematical
and physical concepts described above is 
illusory without caveats or assumptions 
about the underlying physics.
(E.g. ``I believe the underlying mechanism has an
input-output feature that accords with such-and-such  
amplitude-based criterion.)
That this approach is not generally useful
is demonstrated by 
the fact that many linear models can yield arbitrarily
large amplitude flares; for example  
there is no \emph{a priori} limit to the dynamic
range of the model guaranteed,
for any stationary process, 
even if putatively non-linear, 
by the Wold Theorem.

In summary, the concept of non-linearity 
does not transport well to time series in general, 
and astronomical light curves in particular.
Several possible ways to implant linearity 
concepts into astronomical time series analysis 
lack the necessary carefully crafted definitions and 
physical assumptions,
raising the suspicion that 
the term is employed without any
meaning or idea (as is but too frequent).

\subsection{Is \emph{Stationarity} Meaningful for Time Series?}
\label{stationarity}

As with linearity,  
attempts to define
stationarity run afoul of  pragmatic 
difficulties when applied to time series data.
Standard definitions 
invoke \emph{time invariance of statistical quantities}. 
For example constancy of the mean and variance
is termed \emph{weak stationarity}; constancy of
all possible probability distributions 
is \emph{strong stationarity}.  
Many other definitions of stationarity 
are possible, based on invariance of 
various statistical properties.
The concept of \emph{asymptotic stationarity} 
\citep{parzen}
accounts for a system decaying    
away from its initial state, 
approaching 
one of these forms of stationarity in the limit
\cite[see][Sec. 6.2 for a physics setting]{thorne_blandford}.
In non-linear dynamics an important related concept 
is \emph{transient chaos} \citep{young}: 
non-linear pseudo random behavior 
evolving asymptotically to a steady state.


In principle these theoretical concepts
require infinite stretches of data 
to rigorously test for time invariance.
Therefore importing the concepts  
into a realistic data analysis context
requires care. 
A useful concept of stationarity 
must specify how 
independence is to be judged,
the degree of approximation required,
and 
the operative time scale or range.
Furthermore, 
data apparently stationary 
on one time scale 
can easily appear non-stationary 
on another scale, e.g. 
over a longer interval 
 -- and vice versa. 
 In short, the appearance of data within
 the finite window presented by 
 the observations can be misrepresentative of 
 the actual variability,
 both random and systematic.
 Even pulsar rotation,
 one of the best examples of approximate 
astronomical stationarity, 
suffers from non-stationarity 
on long time scales through 
spindown;
in addition short time scale ``glitches''
may or may not be approximately stationary,
 
This issue is related to one raised 
nearly a century ago in 
the classic paper \citep{yule}
dealing with the fact 
``that we sometimes obtain between quantities
varying with the time ... quite high correlations 
to which we cannot attach any physical significance
whatsoever, although under the ordinary test the
correlation would be held to be certainly `significant'."
Yule's cross-correlation problem 
is in a different context,
but his finite-sampling 
and ``cosmic variance'' issues are much the same 
for stationarity.
A common confusion arises from 
random variability on two different time-scales, 
where it is tempting to leap to the view that 
the slower variations are 
nonstationary when judged against 
the faster ones.

The unavoidable conclusion 
that stationarity is ``in the eye of the beholder''
carries with it 
certain ambiguities and subjectivities.
Any assessment of this property 
is dependent on 
(a) the above mentioned 
qualitative and quantitative description 
of method and relevant time scales,
(b) any theoretical or 
other astrophysical considerations
incorporated in the judgement,
and
(c) any preprocessing of the data,
such as removal of trends or intense flares.
Any of these issues 
can dramatically affect conclusions
about stationarity.
The point is that 
the corresponding choices
need to be clearly stated,
not that any one or the other 
is right or wrong.
Furthermore, 
analysis of the data separately 
under both hypotheses (stationary 
and non-stationary) may be productive.
In the end, stationarity is fraught with nearly as
many problematic issues as linearity.

Note that in Section \ref{ar_ma_models} we make use
of the consequences of strict mathematical 
stationarity, and therefore our results must be
used with appropriate caution, with an eye
toward possible effects of non-stationarity.
In this regard 
autoregressive-integrated-moving average 
(ARIMA) models \cite[][e.g.]{scargle_i},
constructed to represent some forms
of non-stationarity, may 
have some application.

\subsection{Is \emph{Multiplicativity} Meaningful for Time Series?}
\label{multiplicativity}

The term \emph{multiplicative} 
\citep[e.g.][]{uttley_mchardy_vaughan}
is similarly
transplanted --
with some of the same issues discussed
above
--
from mathematics to 
physics to time series.
In physics 
the concept posits a 
separation of the system
into subsystems,
each of which contributes a component
to the total flux.
Such a compound process is
termed \emph{additive} or \emph{multiplicative}
depending 
on whether the resulting output is the 
sum or product of that due to 
the components.
Applicability of this
concept 
depends on whether  
separation into 
largely independent 
subsystems is valid, 
and on the correctness 
of the prescription for combining  
their flux contributions.
Justification for this idea 
is sometimes sought from its success 
in other areas of research and from 
log-normal distributions
(log of product = sum of logs = normalcy
as opposed to sum of normals = normal,
both via the central limit theorem).
Here the term additive 
refers mostly to the 
representation of time series 
as the sum of random 
events, as will now be detailed.

\section{The Random Process Models}
\label{ar_ma_models}

A well-known, powerful and flexible model
expresses stationary random processes 
as the output of a linear system
driven by a random input.
This model is known in different fields 
under various names, including:
\emph{autoregressive, 
moving average, 
autoregressive-moving average,
Ornstein-Uhlenbeck,
Brownian motion, 
damped random walk,}
and 
\emph{shot noise}.
Only the discrete forms of
continuous models are of relevance here.
These stationary, linear, and additive models  
are essentially equivalent to each other,
differing mainly in their formal definition and
physical interpretation.
The  
\emph{autoregressive} (AR) 
and \emph{moving average} (MA)
models discussed in this section 
are surrogates for 
members of this class.


The Wold Decomposition Theorem 
\citep{wold}
states that any stationary process 
can be represented in the form of 
the so-called \emph{moving average}
(not the same as \emph{running mean}) 
\begin{equation}
X(n) = \sum_{k} c_{k} R(n-k)  + D(n) \ \ ;
\label{ma_k}
\end{equation}
\noindent
$X(n)$ in the current setting is the flux 
at discrete time $n$
and
$R(n)$ is an uncorrelated random 
process (called the \emph{innovation}).
Stationarity is the only necessary condition;
any other notion, such as linearity, is irrelevant. 
The set of constants 
$\{c_{k}\}$,
called the \emph{moving average coefficients}, 
has two technical properties 
relating to the model flare shape:
\emph{causality} and \emph{minimum-delay}.
This remarkably 
explicit representation 
of the random and non-random aspects 
of an arbitrary stationary process,
and their separation into two  additive terms, 
are among many notable features of the 
Wold Decomposition 
\citep{scargle_findley,scargle_i,brockwell_davis}.
 The component $D$ is a deterministic process,
largely ignored here.
In practice
it is often a constant that can be removed, 
e.g. with the novel
background estimation 
procedure described by \cite{meyer},
or a slowly varying function, 
removable with a detrending procedure.

It is important to realize that Equation (\ref{ma_k}) 
is a theoretical relationship,
asserting the formal equivalence of 
the random process on either side of
the equal sign.
Furthermore, 
there is a family of representations
of a given stationary process
equivalent to each other in this same sense.
For example, 
entirely equivalent to the 
moving average form in 
Equation (\ref{ma_k})
is this \emph{autoregressive}
representation of the same process,
with the same innovation: 
\begin{equation}
X(n) = \sum_{k} a_{k} X(n - k) + R(n) \ \ .
\label{ar_k}
\end{equation}
\noindent
Memory of previous behavior, 
the \emph{Markov property},
is expressed by 
the \emph{autoregressive coefficients} $\{a_{k}\}$.
The term ${a}_{k}X(n-k)$
is the contribution to $X(n)$
of self-memory of the process $k$ steps prior to 
the time $n$.
These forms -- denoted 
AR($K$) or MA($K$), where $K$ is the
number of terms included -- 
are simply 
different ways to represent the same 
random process.
Formally a finite AR(K) process is equivalent to 
MA($\infty$), and 
a finite MA(K) to AR($\infty$);
of course in practice finite approximate versions
are used.
These two representations have different 
astrophysical interpretations,
as we will now see.

It is quite natural to picture 
the moving average as modeling
the observed flux $X(n)$ at time $n$ as 
the superposition of randomly 
occurring \emph{flares}
(also called\emph{ pulses, events, shots, filters}, etc., depending on context),
for which the term 
\emph{shot-noise} is often used.
The flare shape is 
determined by the coefficients \{$c_{k}$\}.
The flare amplitude at time $n$ is $R(n)$.
See \citep{scargle_findley,scargle_i}
for discussion of the role of 
independently distributed innovations,
corresponding, e.g., to the assumption
that the light curve is generated by 
physically separated subsystems
not communicating with each other,
whose outputs are therefore statistically independent.
(If, on the other hand, $R$ is normally distributed
the above model 
is  a \emph{Gauss-Markov process},
of no interest here for reasons outlined below.)

\begin{figure}[htb]
\includegraphics[scale=.43]{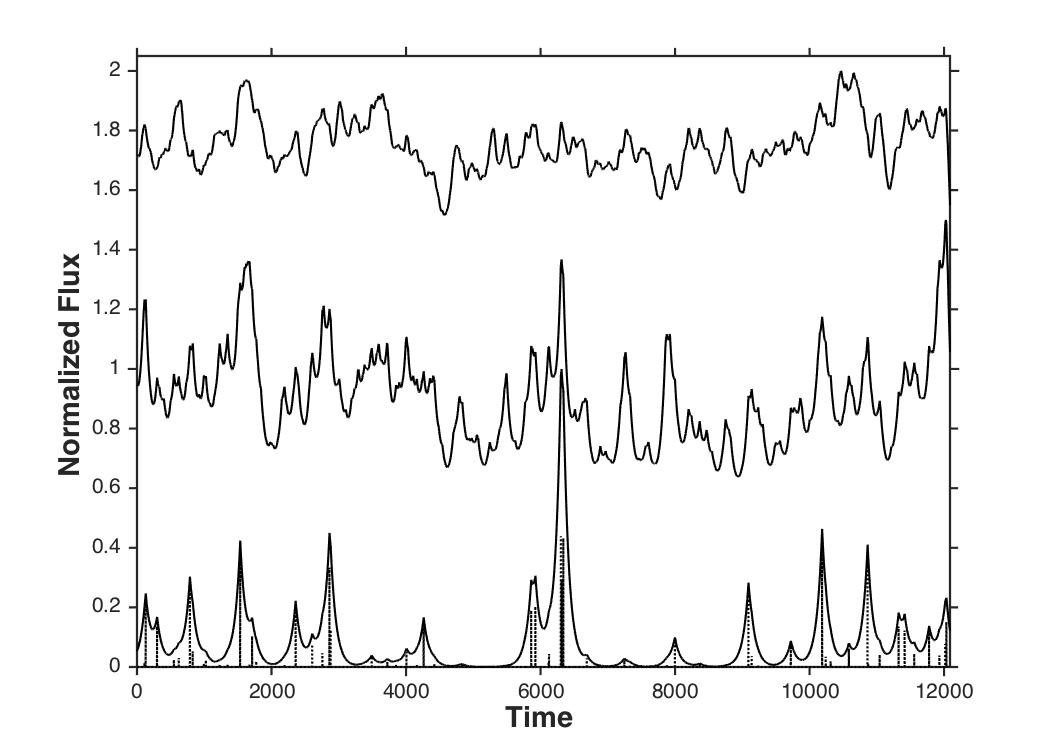}
\caption{Simulated moving average light curves:
Flare shape  $C \sim$e$^{- | t - t_{0} | / 81.3 }$,
a symmetric two-sided exponential;
innovations $R=U^{\alpha}$; $U$ 
uniformly distributed random numbers;
$\alpha$ = 10 (top), 
100 (middle), 
and for 1000 (bottom) U$^{\alpha}$ 
is also shown with dotted lines.
By favoring large values 
and suppressing small ones,
increasing $\alpha$ 
yields increasingly sparse
innovations $U^{\alpha}$.
}
\label{fig_0}
\end{figure}

Figure (\ref{fig_0}) depicts three simulations of Eq. 
(\ref{ma_k}) differing only in the distribution 
of the innovation.
For the sparsest case 
(the bottom curve) 
the events are relatively isolated 
from each other.
For the middle curve the innovation 
is less sparse, so 
there is considerable overlap 
between events.
The top curve
approaches the Gaussian limit 
where the true flare shape
cannot be determined by any algorithm,
because the high degree of  overlap hides 
all information beyond second order statistics.

The autoregressive version of the Wold Representation,
Equation (\ref{ar_k})
has the form of a linear system driven by a random
input.
While less visual, this interpretation 
can in principle be tied to a physical 
picture of an underlying dynamical 
process, with short- or long-term memory
depending on the number of terms included.
However one should keep in mind
that the two models are equivalent, 
interchangeable representations.
With a slightly different notation 
than adopted here,
the AR and MA coefficients 
are convolutional inverses
of each other.
The innovation can be determined 
by convolving the time series data 
with an estimate of the AR coefficients.
Then the modeled data,
defined by convolving this innovation
with the MA coeffecients,
exactly reproduce the light curve.
For more details see 
\citep{scargle_i}
or statistics textbooks 
\citep{priestley,brockwell_davis}.


A few more comments on the 
significance of these models
are in order.
In both MA and AR formulations 
the innovation 
encodes the amplitudes of flare-like
events.  
Its statistical distribution,
not known a priori, 
is a key goal  
of data analysis.
For astronomical light curves
positive-definite 
innovations are relevant, 
because fluxes are non-negative.
Gaussian innovations 
have the problem of negative values, as well 
the fact that the degeneracy 
among mixed-delay and mixed-causality 
models cannot be resolved 
for a Gaussian process 
\emph{by any algorithm} \citep{scargle_findley,scargle_i}.
Crucially, then,
\emph{astronomical time series data 
need to be modeled with 
positive definite, non-Gaussian 
innovations.}

The \emph{ Ornstein-Uhlenbeck 
process} (OU),
of importance in mathematical physics
and increasingly invoked in 
astrophysics \citep{kelly_2009,kelly_2011,takata,kelly_2}
is a continuous version of Equation (\ref{ar_k})
with only the $k=1$ term.
It is 
usually defined as the solution of 
stochastic differential equations such as 
the Langevin equation, the 
Fokker-Planck equation,
or a continuous version 
of  AR(1) known as  the Vasicek model 
in econometrics.
The terms 
\emph{Brownian motion},
\emph{damped random walk},
and  \emph{L\'evy process}
are also used for essentially the same model.
See 
\cite{kelly_2009}
for application to quasar light curves
and 
\cite{kelly_2011} 
for further details.
What is in common to all of these formalisms 
are the same properties described above:
superposition of randomly occurring events
(explicit in MA) and memory of the past (explicit in AR).

\section{The \emph{RMS-Mean Flux} Relation}
\label{rms}

The vaunted rms-mean flux relation 
explores possible dependence of 
flux variability on the flux itself.
It is based on straightforward computation of 
the mean and standard deviation 
within subintervals of 
the total observation span.
Free choices include 
the manner of correcting for 
observational noise (discussed below), 
the length and possible overlap of the subintervals,
the binning employed in smoothing scatter plots, 
and possible data selection 
in the time or frequency domain.
While \cite{uttley_4} 
provide complete detail and Python code,
some others do not
give enough  information for one to 
reproduce the results.

The frequently observed rms-mean flux behavior 
as described in Section \ref{previous}  
could derive from a universal physical process 
(e.g. turbulent accretion or
jet dynamics), 
or from generic statistical  
properties of light-curves 
(e.g. stationarity or additivity).
Evidence  favoring the latter
is next provided by
demonstrating that AR and MA models,
which are not based on 
specific physical processes,
adequately 
reproduce the 
observed relations.

\begin{figure}[htb]
\includegraphics[scale=.47]{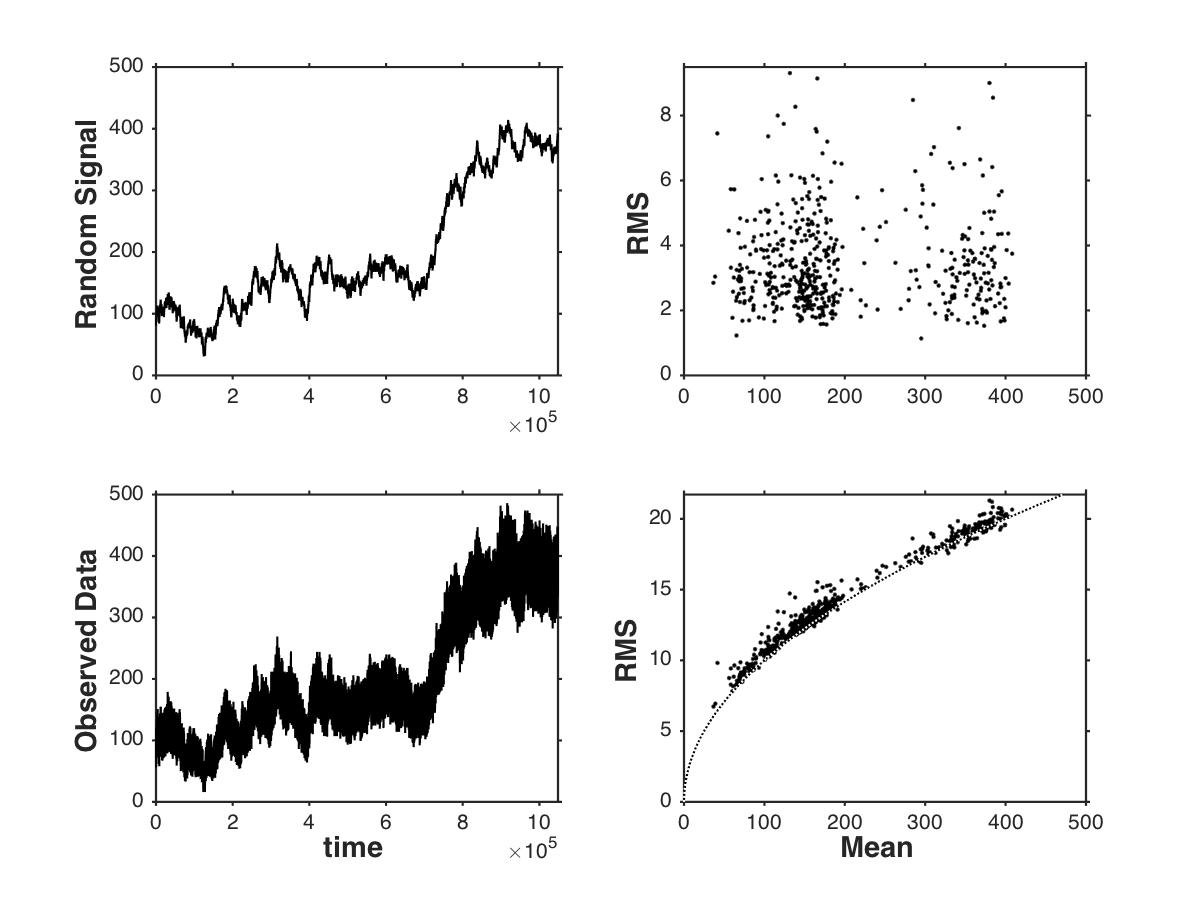}
\advance\leftskip-1.5cm
\caption{A simulated random walk time series.
Upper-Left: cumulative sum of 
slightly biased normally distributed variables,
renormalized.
Lower-Left: 
The same data 
processed by applying the 
Poisson operator.
Right Panels: scatter plots of 
root-mean-square flux vs. mean flux,
derived from the 
corresponding data on the left-hand side.
The solid line in the bottom-right
panel is the theoretical  square-root relation 
for the Poisson distribution.
This figure demonstrates 
the effects of photon counting fluctuations, 
but for the most part 
these are accounted for in the data processing.
}
\label{fig_poiss}
\end{figure}

First note that 
observed fluxes based on photon counts
have a built-in linear dependence of 
flux variance on mean flux
(and hence a square-root dependence
for the standard deviation)
directly through 
photon counting fluctuations.
Figure \ref{fig_poiss} 
demonstrates this fact
by comparing a synthetic 
random light curve against
a sampled version representing 
the additional variance 
due to photon count 
fluctuations.\footnote{This figure accentuates 
that light curves almost always 
embody 
two independent 
random processes:
intrinsic variations (signal) and 
photon fluctuations (noise).
This 
is known 
as a 
 \emph{doubly stochastic},
 or \emph{Cox process}.}
 Such sampling is 
neither additive nor multiplicative,
rather the result of applying
an operator to the light curve 
to simulate 
observations that obey 
the Poisson
distribution with mean photon
rate equal to the source flux 
at each moment of time. 
As seen in the bottom right-hand
panel these samples 
show the expected square-root 
rms dependence, 
shown as a line,
easily mistakable for a linear
relation over much of its extent.

This effect is probably not
directly responsible for most of the
reported rms-flux
relationships.
\cite{uttley_4} provide a link to python code,
which includes a correction 
subtracting the 
``... expected contribution of 
the observational error 
to the total variance ... to give
the intrinsic variance ... ''
Implementing this correction here,  by 
taking such surrogate observational 
errors  to be Poisson 
fluctuations, 
largely removes the dependence shown in 
the lower-right-hand panel of 
Figure \ref{fig_poiss}.
Nonetheless it is reasonable that 
imperfect estimation of, or accounting for, 
this photon counting contribution to the variance 
might affect the derived relation.
It appears that some studies may 
have not carefully distinguished 
between observational and source-intrinsic
variance in making this correction,
or possibly not made the correction at all.

\clearpage

The relevant quantities are 
the mean and standard deviation 
of the flux, averaged over finite
subintervals of the time series.
Evaluating in this sense 
the expectation $E$
of the equation for the second order 
autoregressive model AR(2), i.e. 
\begin{equation}
X(n) = a X(n - 1) + b X(n-2)+ R(n) \ ,
\label{ar_2}
\end{equation}
\noindent
yields a linear relation between 
the flux and innovation means:
\begin{equation}
E(X) = {1 \over 1-a - b} E(R) \ .
\end{equation}
\noindent 
The same relation 
follows from the moving average representation 
-- cf. Eq. (\ref{ma_k_1}) below -- 
yielding the intuitively expected result
that the proportionality constant 
${1 \over 1-a - b}  = \sum c_{k}$,
the total area of the flare shape.
To estimate the variance 
we need to find the mean of
\begin{equation}
  X^ 2(n) = a^2 X^2(n-1) + b^2 X^2(n-2) +  R^ 2(n) \\
 + 2 a b X(n-1) X(n-2)  
 +  2 a X(n-1) R(n) 
 +  2 b X(n-2) R(n) 
\end{equation}
\noindent
Since previous values 
of $X$ are independent of 
the current value of the innovation,
we have
\begin{eqnarray}
E( X^ 2 ) = a ^2 E( X^2  )+ b ^2 E( X^2  ) +  E(R ^ 2 ) 
+ 2 ab \  \rho(1)  
+ 2 ( a + b ) E(X) E(R) \ , 
\end{eqnarray}
\noindent
yielding 
\begin{equation}
E( X^ 2 ) = 
{E(R ^ 2 ) 
+ 2 ab \  \rho(1)  
+ 2 ( a + b ) (1 - a - b) E^{2}(X) \over ( 1 - a^2 - b^2  ) } \ .
\end{equation}
\noindent
The variance $E( X^ 2 ) - E^ 2( X )$  is thus 
\begin{equation}
\sigma_X ^{2}  
=  \alpha E^{2}(X) +  \beta \ ,
\label{std_ar_2}
\end{equation}
\noindent
where 
\begin{equation}
\alpha =  {  2 ( a + b ) ( 1 - a - b ) \over ( 1 - a^2 - b^2  )  }
\end{equation}
\begin{equation}
\beta =   { 2 a^{2}b \  \rho(0) 
 \over ( 1 - a^2 - b^2  ) ( 1 - b )  } 
+ E(R^{2}) {1  \over  1 - a^2 - b^2  } 
- E^{2}(R) {1  \over  (1 - a - b ) ^2 } \ , 
\end{equation}
\noindent
using the Yule-Walker solution 
$\rho(1) =  a \rho(0) / (1- b)$
\citep{priestley}.
For AR(1) simply set $b=0$:
\begin{equation}
\alpha =  {  2 a  \over ( 1 + a   )  }
; \hskip 0.5in \beta =  
 E(R^{2}) {1  \over  1 - a^2   } 
- E^{2}(R) {1  \over  (1 - a ) ^2 } \ , 
\end{equation}

Similar formulas can be obtained for the
moving average representation.
The expectation value of Eq. (\ref{ma_k}) gives
\begin{equation}
E(X) = E(R) \sum_{k} c_{k} \ ,
\label{ma_k_1}
\end{equation}
\noindent
stating that the average output 
is the product of the area of the flare 
profile $C = \{ c_{k} \}$,
and the mean innovation.
A further consequence of  Eq. (\ref{ma_k}) is 
\begin{equation}
E(X^{2}) = \sum_{j} \sum_{k}  c_{j} c_{k} E[ R(n-j)  R(n-k) ] \ \ ,
\end{equation}
\noindent
giving, 
for the sum of the diagonal and off-diagonal terms,
respectively:
\begin{equation}
E(X^{2}) =  E(R^{2})  \sum_{j} c^{2}_{j}  + 
E^{2}(R) \sum_{j \ne k} c_{j} c_{k}  \ ,
\end{equation}
\noindent
yielding
\begin{equation}
\sigma^{2}_{X}  =  E(R^{2})  \sum_{j} c^{2}_{j}  + 
E^{2}(R) \sum_{j \ne k} c_{j} c_{k} -
E^{2}(R) \sum_{j} \sum_{k}  c_{j}  c_{k} \ ,
\end{equation}
\noindent
or
\begin{equation}
\sigma^{2}_{X}  
= \sigma^{2}_{R}  \sum_{j} c^{2}_{j} \ .
\label{ma_rms_flux}
\end{equation}
\noindent
This tidy formula seems very different 
from Eq. (\ref{std_ar_2}) but is, in fact,
equivalent to it.
For the two relevant 
autoregressive models, this can be seen 
with a little algebra, 
e.g. using  for AR(1) the sum 
$\sum_{j} c^{2}_{j}  = {1 \over 1 - a^{2}}$.
While the two simplified expressions for variance 
as a function of the model parameters
and innovation,
are suggestive of a linear RMS-mean
relation -- 
Equation (\ref{std_ar_2}) 
if $\beta$ is small,
and 
Equation (\ref{ma_rms_flux})
if $\sqrt{ \sigma^{2}_{R}  \sum_{j} c^{2}_{j} }  \sim E(X) $ --
the actual dependence on mean flux 
is implicit, not explicit.

\begin{figure}[htb]
\advance\leftskip-2cm
\includegraphics[scale=.5]{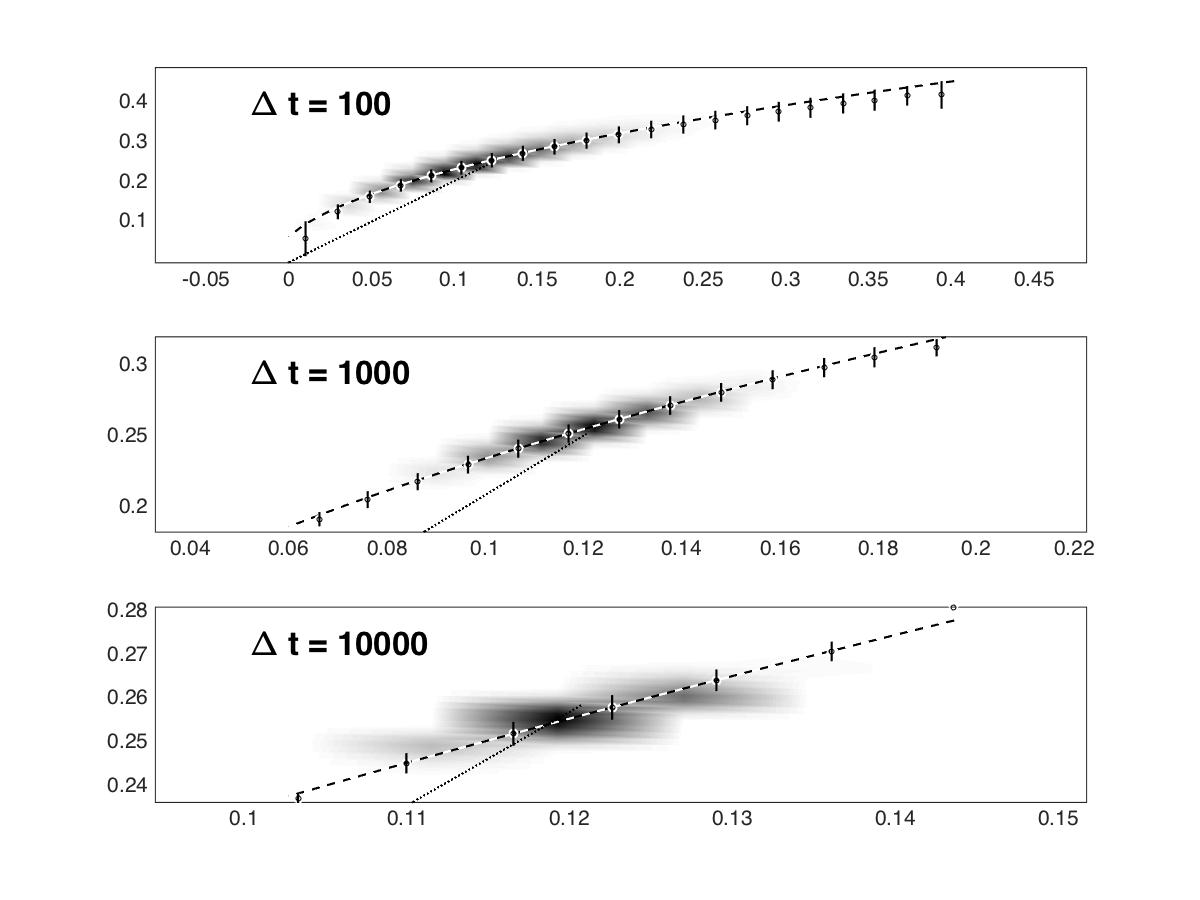}
\caption{Grey scale plots of the density of 
points in the 
rms (standard deviation)  vs. mean flux plane,
for simulated AR(1) light curve:
$a= 0.5$, 
$R$ distributed as $U(4)$,
i.e. uniform random numbers on $(0,1)$ 
raised to the power $4$,
and truncated below 0.75.
The total number of points
generated is $N = 2^{30} = 1,073,741,825$.
(No simulated observational errors were 
applied to these data.)
Three different
sizes of the time interval
over which the averages were
computed are indicated at the
top-left of each panel;
the corresponding number 
of points entering the density 
is $N$ divided by this number.
The dashed lines are least-squares 
fits to Eq. (\ref{std_ar_2})
evaluated at these mean flux values.
The dotted lines originate at (0,0)
to indicate that the quasi-linear relations
do not intersect the origin.
The points and error bars 
are means and standard deviations
averaged over a set of evenly spaced intervals.}
\label{sim_rms_ar1}
\end{figure}

If the distribution of $R$ is known, 
we can evaluate $\sqrt{ \sigma^{2}_{R}}$.
In the case of a power law distribution of the 
form\footnote{Not to be confused with the distribution
obtained by raising uniformly distributed
random numbers to a power, as 
used in the construction of Fig. \ref{sim_rms_ar1}.}
\begin{eqnarray}
F(R) =  F_{0} \ R ^{\alpha}  \ \  \mbox{for} \ \ 0 \le R \le R_{0}  \hskip0.1in (0 \ \ \mbox{otherwise})
\end{eqnarray}
it is straightforward to find the normalization factor
\begin{equation}
F_{0} = ( \alpha + 1)  R_{0}^{-(\alpha + 1)}
\end{equation}
and the first and second moments 
\begin{equation}
E(R) = 
F_{0} R_{0}^{\alpha + 2} 
( \alpha + 2) ^{-1}\ ; \ \ \ \ 
E(R^{2} ) = F_{0}  R_{0}^{\alpha + 3} 
(\alpha + 3)^{-1} \ , 
\end{equation}
\noindent
and with a bit of algebra 










\begin{equation}
\sigma^{2}_{R} =
(\alpha + 1) 
 ( \alpha + 2 ) ^{-2} 
 (  \alpha + 3 )^{-1}
R_{0}^{2} 
\end{equation}
Using Equation (\ref{ma_k_1})
to put this relation 
in a more easily interpretable form,
we find







\begin{equation}
\sigma^{2}_{X} =
(\alpha + 1 )^{-1} (\alpha + 3 )^{-1} 
{\sum c_{k} ^{2}  \over
( \sum_{k} c_{k} )^{2} } \ 
E^{2}(X) \ , 
\end{equation}
\noindent
corresponding to a linear rms-mean flux relation.

Turn now to numerical simulations.
Figure \ref{sim_rms_ar1} 
shows plots of $\sigma_{X}$  vs. E$(X)$ 
for the sample AR(1) process
defined in the caption.
The data used to make the figure
are large sets
of pairs of values,
means and standard deviations, 
evaluated over non-overlapping sub-intervals.
To avoid over-plotting that would 
confuse a simple scatter plot,
this figure shows grey-scale
representations of point density.
In addition, 
for comparison with most published
figures, points (show as circles)
 and error bars 
averaged over  flux bins
are displayed.
The three panels 
cover a range of two orders of magnitude
in sub-interval length.
The dashed lines are fits 
to points generated by evaluating
the square root of
Equation (\ref{std_ar_2}) 
at the corresponding 
values of the abscissa.

For this particular model
the figure demonstrates linear or slightly curved 
dependence of 
rms on mean flux.
A small study of other model
orders, parameters
and innovations suggests that 
this result is characteristic of the class
of autoregressive/moving average models,
with the shape of the relation being 
determined by the distribution of the 
innovation as suggested by 
Equation (\ref{ma_rms_flux}).
Importantly the non-linear rms-flux 
relations discussed 
by \cite{alston_et_al} 
and 
\cite{alston} 
are curved in the same sense as 
in the first two panels of Figure \ref{sim_rms_ar1}.
Some published scatter plots 
seem consistent with either linear or quadratic 
forms, within statistical uncertainties.

\begin{figure}[htb]
\includegraphics[scale=.425]{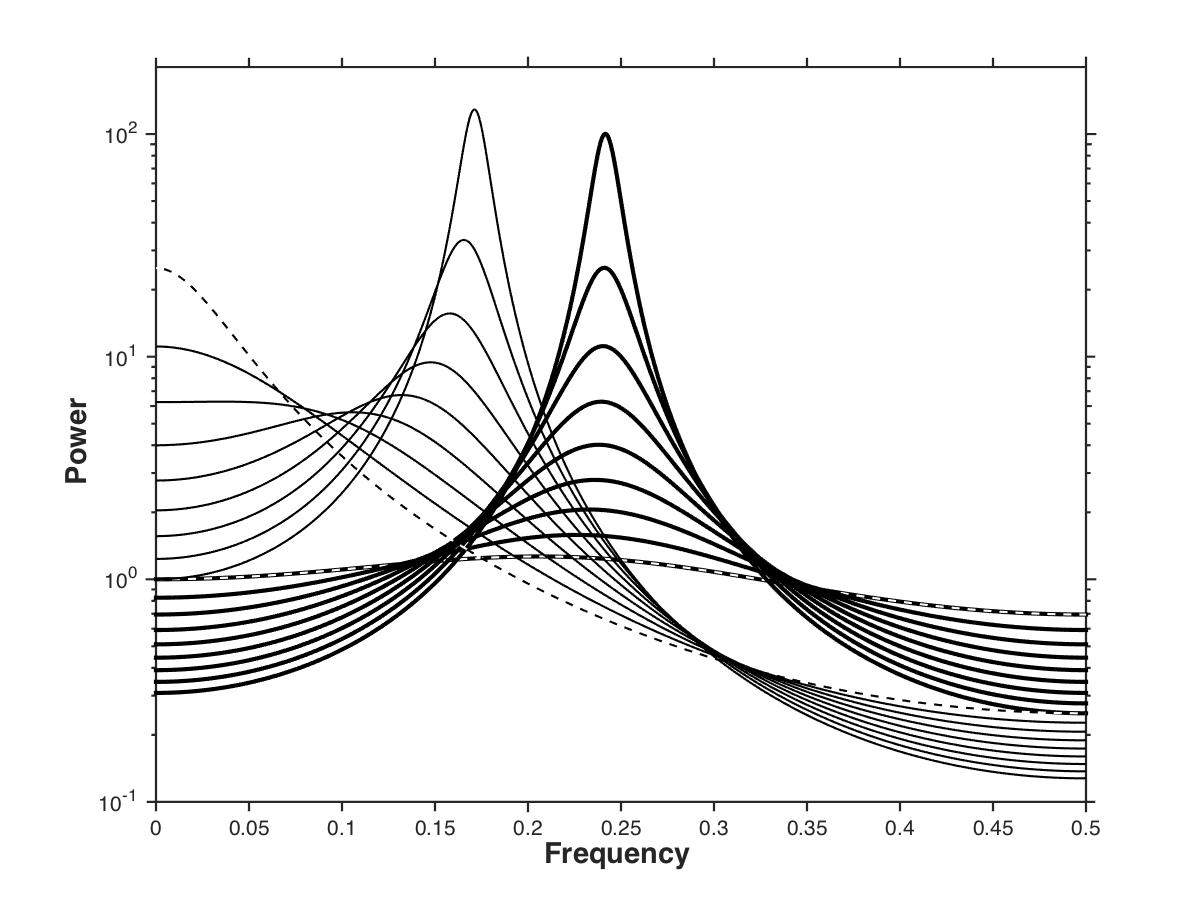}
\caption{Logarithms of AR(2) power spectra
from
Equation (\ref{ar_2_pow});
$a$ = -.1 (thick lines) and $a$ = -.9 (thin lines).
In both cases $b$ ranges from -0.1 to -0.9
in steps of 0.1, with  $b = -0.1$ a dashed line.
}
\label{ar_2_power}
\end{figure}

By the way,
the textbook power spectrum of 
the AR(2) process is 
\begin{equation}
P(f) = { \sigma_{X}^{2} \over 1 + a^2 + b^2 
+ 2 a ( 1 - b ) \mbox{cos}( 2 \pi f ) 
 - 2b \mbox{cos}( 4 \pi f ) }   \  \ .
\label{ar_2_pow}
\end{equation}
\noindent
The wide range of shapes yielded by this formula,
samples of which are depicted in Figure \ref{ar_2_power},
is perhaps not widely appreciated.
Section 3.5.3 of \citep{priestley} 
details the more intricate
formula for the autocorrelation
function.
It is noteworthy that  time series
generated by this elementary model 
can display both random 
and quasi-periodic (Priestley uses the term 
\emph{pseudo-periodic}) behavior, 
as suggested by Figure \ref{ar_2_power}.

\clearpage

\section{Flux Distributions}
\label{flux_distributions}

Wide interest in 
the distribution of 
measured flux values 
largely focuses on the
binary choice between normal
and log-normal \citep{log_normal_book}.
That this may be a false choice
can be seen from 
the following computation 
of the exact distribution
for an arbitrary autoregressive 
process.

A straightforward evaluation of 
the distribution $P_{X}(X)$ of $X$, 
in terms of the distribution 
$P_{R}(R)$  of the innovation $R$,
starts from the  moving average representation in
Eq. (\ref{ma_k}).
This equation holds for 
causal ($k>= 0$), 
acausal ($k<=0$)
and mixed representations 
($k$ unconstrained).
We invoke two text-book results:
(1) The distribution of the sum of
two random variables is the 
convolution of their distributions.
(2) The distribution of a constant $C$ times 
a random variable $R$ is
$P_{CR} ( C R ) ={1 \over C} P_{R}( {R \over C})$.
With these facts Equation (\ref{ma_k}) yields:
\begin{equation}
P_{X}(X)= \prod_{k} F_{c_{k} R}( c_{k} R ) = 
\prod_{k} {1 \over c_{k} } F_{R}( {R \over c_{k} } ) \ ,
\label{ma_dist}
\end{equation}
\noindent
with $\prod$ denoting the convolution operation.

\begin{figure}[htb]
\includegraphics[scale=.45]{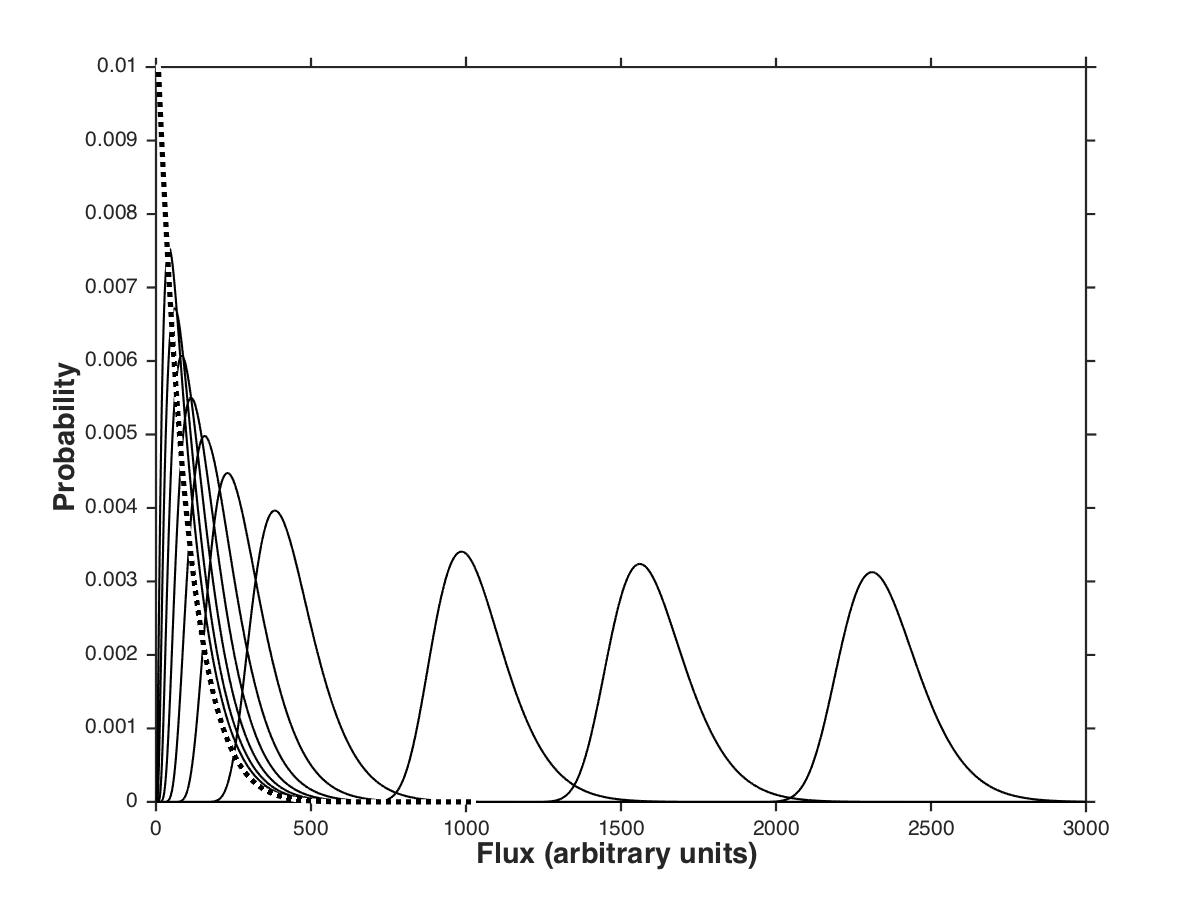}
\caption{Flux distributions 
from AR(1) proceesses.  
The innovation 
is a power law distribution -- $U^{10}$ 
in the notation in \cite{scargle_i}, i.e. a uniform
distribution raised to the $10$th power --
but reversed 
so that the probability decreases with increasing 
flux amplitude, and is plotted as a dotted line.
The other curves are the flux distributions 
obtained from Eq. (\ref{ma_dist})
for a sequence of autoregressive 
parameters $a$ from the set 
$( .2 \  .3 \  .4 \  .5  \ .6 \ .7 \ .8 \  .9 \ .925 \ .94 )$;
the peaks of these distributions  
appear ordered in the same sense as these 
parameter values, increasing to the right.
}
\label{flux_dist_ar1}
\end{figure}

This formula is exact for 
an arbitrary moving average process.
Figure \ref{flux_dist_ar1} depicts 
these distributions for the special 
case of a first order autoregressive 
process with coefficient $a$.
A monotonically decreasing 
power-law was chosen for the 
innovation distribution $P(R)$.
For small values of $a$ -- almost no memory
of previous values -- this output process is 
a nearly unaltered version of the input
innovation, so the distribution is 
close to that of the innovation itself 
(shown as a thick dotted line).
As $a$ increases the distribution
broadens,
for a while 
maintaining the 
high-end tail lending 
the appearance of  log-normalcy.
However, as $a$ approaches $1$, 
corresponding to a very  strong memory,
the distribution approaches a symmetric
normal form; this is completely 
understandable through the
central limit theorem and the fact that 
as $a \rightarrow 1$   
many random variables are added
together via equation (\ref{ma_k}).
Of course this distribution 
must have zero weight for 
negative fluxes and cannot be
exactly Gaussian.

In summary: the shape of the flux 
distribution for this linear process
depends on two things: the distribution 
of the innovation and the value of the
decay constant $a$. 
With a toy but not unrealistic power
law distribution of input flare amplitudes
(the innovation), 
distributions resembling normal 
or log-normal ones can be reproduced.
Assertions that non-linear or ``multiplicative'' 
dynamical processes necessarily underly
astrophysical systems based on log-normalcy
are thus disproved.

\clearpage

Furthermore, 
the logarithm is not the only relevant transform,
and generally speaking is not particularly 
suited for making distributions Gaussian.
\cite{box_cox} is a classic study 
of normalizing transformations
in the form of simple power laws,
as describe in Fig. \ref{transform} below.
In his 1981 Wald Memorial Lecture \cite{efron}
derived conditions under which 
distributions can be normalized by 
monotonic transformations,
exhibited formulas for calculating them,
and elucidated the relationship
between normalization and variance stabilization.
Based on work of 
\cite{curtiss}, \cite{barlev_enis}
derived explicit formulas for several 
\emph{variance stabilizing transforms}
including 
\begin{equation}
\mathcal{A}_{\alpha,\beta} (X)  =
( X + 2 \alpha - \beta ) ( X + \alpha )^{-1/2} \ .
\label{bar_lev_enis}
\end{equation}
To construct Figure \ref{transform} 
we optimize the normalcy yielded by 
this form, with respect to its parameters
rather than use formulas -- 
like the well known Anscomb transform
$2 \sqrt{ X + 3/8 })$ -- 
optimal for assumptions possibly 
not applicable to these data.
This figure 
displays distributions of 
the Cyg X-1 flux values
analyzed by \cite{uttley_mchardy_vaughan},
helpfully provided in a link to a Python Jupyter
notebook by \cite{uttley_4}, 
both in raw form and as 
transformed by the logarithm and
two other functions.
(These authors discussed 
variance stabilizing transforms in 
a related context.)
The middle panel 
is for the logarithm and 
optimized power-law 
$F \rightarrow F^{a}$
(optimized to yield the 
minimum rms-residuals
from Gaussianity),
for which the size and distribution
of the residuals are 
essentially the same.
The right-hand panel 
shows the distribution
yielded by the transform 
in Eq. (\ref{bar_lev_enis}) 
optimized with respect to 
$\alpha$ and $\beta$.
Note that here residuals are smaller 
and more randomly distributed 
than for the logarithmic or power law
cases.
At least in this anecdotal case log-normalcy is not magical. 
A number of statistical 
procedures, 
e.g. 
Kolmogorov-Smirnov,
Kuiper,
Shapiro-Wilk 
and
Jarque-Bera tests -- 
with careful attention to 
associated caveats and assumptions --
can be used to formally assess
goodness-of-fit of data to a given distribution.

\begin{figure}[htb]
\advance\leftskip-1.5cm
\includegraphics[scale=.45]{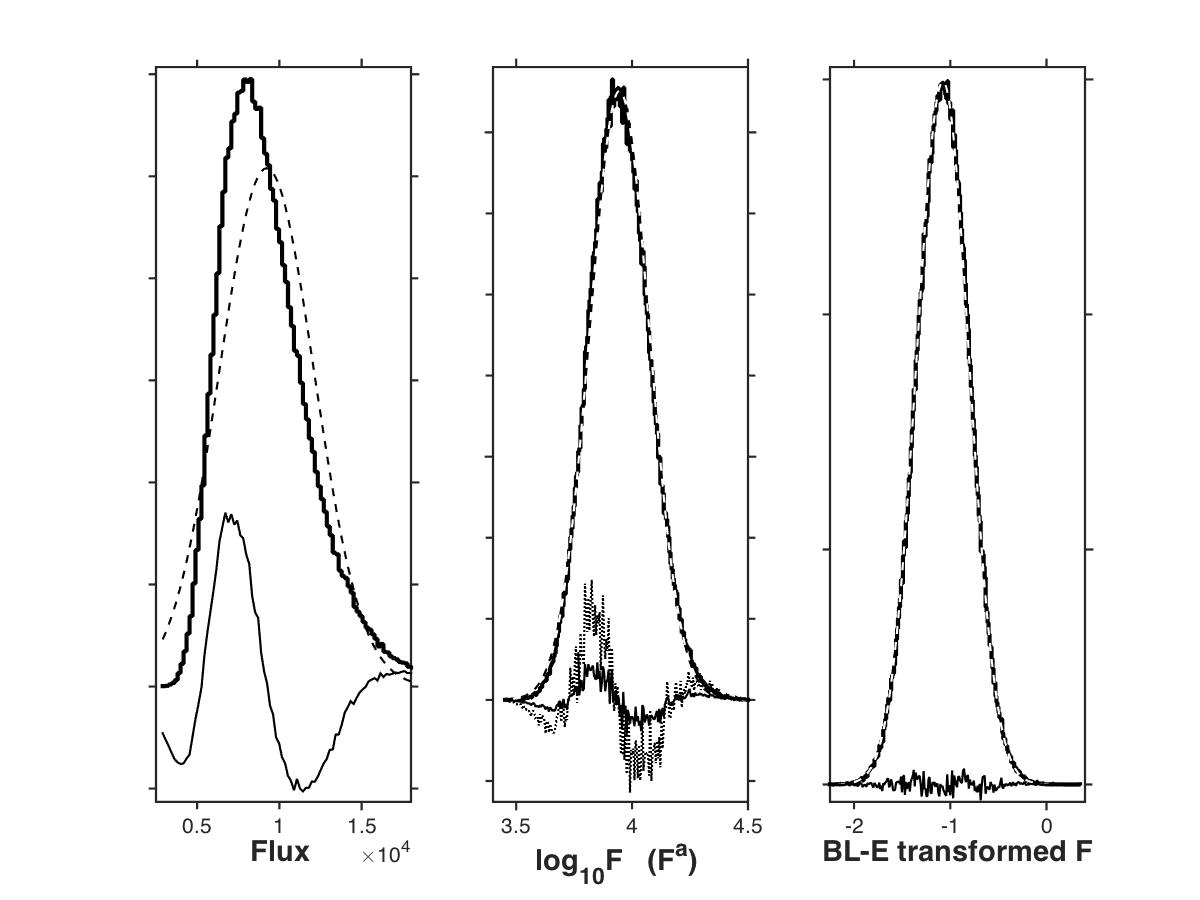}
\caption{Cyg X-1 flux distributions 
(thick solid lines), best fitting normal functions 
(dashed lines) and residuals (thin lines).
Left: raw fluxes.
Middle: log and power law transformed 
fluxes (optimum $a=0.04$).
For the latter 
the residuals (dotted line) 
are multiplied by 3 for clarity,
but are essentially indistinguishable
from those for the log transform. 
Right: Optimized Bar-Lev and Enis transform,
with $\alpha = 0.01$ and $\beta = 0.664$.
}
\label{transform}
\end{figure}
\clearpage

\section{Discussion}
\label{discussion}

For several reasons 
the class of autoregressive/moving average 
and related processes 
discussed here 
form a powerful and flexible set of models 
for time series data observed in  
a variety of flickering astronomical sources.

{\bf Astrophysical realism:}
The curves plotted in Figure \ref{fig_0}
are visually similar to 
time series data for 
many flickering astronomical sources,
for example the gamma-ray 
light curves 
of the active galactic nuclei discussed by
\cite{meyer}.
Even just these three samples 
suggest that the wide range 
of intermittency 
in the observations
can be represented with innovations
of varying degrees of sparsity.
The degree of flare overlap,
ranging from isolated discrete events 
to very considerable merging 
of the profiles of successive flares,
is controlled by the distribution of the innovation.
For modeling of actual time series data, 
if deterministic background trends,
observational errors,
and optimized parameters
(flare shape and innovation) are incorporated, 
these sample simulations become 
even more realistic.

{\bf Implementation properties:}
The Wold Theorem and its constructive proof 
\citep{scargle_i}
both guarantee the existence of these
linear and 
remarkably specific models for arbitrary 
stationary data, and
provide a pathway to estimating them.
Their second order statistics 
(power spectra and autocorrelation functions)
are those of the event shape $C$, 
and therefore are essentially arbitrary -- flexible 
enough to match the
second order statistics 
of any time series data.
The results derived here 
demonstrate that they can 
match flux distributions and  
rms-mean flux relations as well.
The models fit the data exactly,
so quality-of-fit resides in the
statistical properties of the innovation.
A great many theoretical properties 
of this class of random 
processes have been explored 
and supported by efficient algorithms,
included in standard data analysis
systems.

{\bf Physical Interpretation:} 
The models have natural and useful interpretations.
The moving average form embodies 
a random sequence of flares of various amplitudes.
Flare shape and amplitude properties 
are separated, 
each leading 
to useful comparisons with physical theory. 
The autoregressive form explicates 
the standard short-term or long-term 
memory characteristics of  Markov processes
and random walks.

\emph{Are these models  linear?}
Well, that depends on what you mean!
Equation (\ref{ar_k})  has 
a clear linear input-output structure: 
if the innovation is a linear combination
of two or more independent innovations, 
the output is a linear combination of 
the corresponding outputs.
In addition, this equation
represents memory -- the Markov property --
as a linear combination of prior values.
\emph{The moving average/autoregressive model
is linear in both of these senses.}


A comprehensive development 
of analysis tools in this framework 
is left  to another publication.
The main purpose here is to 
clarify some methodological issues 
relevant to deriving physical properties 
of astronomical sources 
from statistical properties of their
time series data.
To wit: the following statistical properties 
can be derived from time series data:
power spectra,
phase spectra, 
rms-mean flux relations, 
flux distributions,
measures of stationarity or 
intermittency, 
and dynamic range.
In many publications 
these properties,
separately or in combination,
have been taken to 
imply that the observed system must 
(or must not) have, 
separately or in combination: 
non-linear dynamics, 
non-stationary time evolution, 
multiplicative component processes, 
or representability in terms of random flares
(``shot noise'').
The models discussed here 
serve as counterexamples
to such assertions.
Without any such physical properties 
these stationary, additive, 
linear random process models 
can generally speaking match the observations,
e.g. by displaying linear dependence 
of flux standard deviation on flux mean 
and log-normal flux distribution.

This conclusion supports the view that 
the near ubiquity of relevant properties 
of light curves 
are likely statistical features
that are generic, 
at least within 
the arena of astronomical time series, 
not necessarily associated with 
a universal physical mechanism.
Specifically, 
mere stationarity is a sufficient condition 
for the existence of the relevant linear 
models, which have none of the exotic
properties listed above.


Examples of the mistaken implications 
mentioned above are too many to 
cite in detail.  
A common assertion is that a linear
rms-mean flux relation in the time series data
implies non-linearity or non-stationarity 
of the physical process underlying the 
observed flickering.
Also disproved are similar 
assertions  about ``shot noise,'' 
such as that it must have,
or must not have, certain features.
In some cases this may be a semantic
issue; if shot-noise is defined to have
a stationary innovation, then it obviously 
cannot be non-stationary.
But the flexibility of the distribution
of the innovation allows either behavior.
Assertions of the form
\begin{quote}
Given statistical property X,
the underlying process must have
property Y and cannot have property Z.
\end{quote}
generally speaking 
should rather be phrased
\begin{quote}
Given statistical property X,
the underlying process may have
property Y
(but there are simple models that do not have this
property),
evidence for which could
be obtained from other considerations
or other data.  Regarding Z: Never say never.
\end{quote}

What approaches can avoid the missteps
cautioned against here? 
Careful attention to 
the definitions of the various 
time series statistics, and 
clearheaded evaluation of the astrophysical 
consequences of the corresponding observations, 
are obviously needed.
Sophisticated or detailed 
models should be 
evaluated against the 
null hypothesis that simple 
(linear, additive and stationary)
autoregressive/moving average 
models adequately represent the observations.

These models can provide direct information about 
variability duty cycles and the statistical 
distributions of flare amplitudes 
(through the innovation) and 
shapes 
(through the model coefficients).
In addition,  the formalism 
described here holds the promise of 
deriving physically meaningful 
properties of the underlying Markov process
via values of $\alpha$  and $\beta$ 
in Equation (\ref{std_ar_2}).
read off the 
RMS-mean flux relation.
Toward this end, 
linkage of properties of the
innovation and flare shapes 
to astrophysical characteristics
would be useful.
A specific example:
asymmetric flares, such as fast rise 
and exponential decay (FRED),
might indicate 
explosive injections followed by 
expansion and cooling,
or delays across a curved 
relativistic jet front \citep{fenimore};
symmetric flares might 
point toward jets  
randomly sweeping by the line of sight
to the observer \citep{nemiroff}.

All of these approaches can profit 
from an openness to what 
generic physical characteristics 
are indeed implied by the observations,
notwithstanding the cautions urged 
in this discussion.
For example some of the conclusions
that do not follow ineluctably from the statistical 
characteristics discussed in this note
might be supported by auxiliary 
information -- notably 
time resolved measurements of 
polarization  and energy spectra.

Lastly, the model framework
discussed here is far from the final word.
Not all of issues affecting  practical use of 
this class of models 
have been resolved.
While the ambiguities related to causality 
and delay properties have been 
addressed via a generalization of the
Wold Decomposition
\cite{scargle_findley,scargle_i}, 
there remains  the obvious difficulty 
of flare shapes that depend on time,
either systematically or randomly.
\cite{press} discussed 
\emph{scale superimposition processes} --
moving averages 
in which the flare shape is 
stretched by a randomly varying factor.
If the stretch process is stationary, 
the Wold Representation 
expresses the resulting time series 
as the superposition of a single fixed shape, 
which must somehow be an average 
of the stretched ones.
The details of how this 
all works are not obvious.

\section{Acknowledgements}
\label{cknowledgements}
Thanks especially to 
Sarah Wagner and Jay Norris for
many useful for comments,
and also to 
Krista Lynne Smith, Roger Romani, 
Paul Burd, 
Aneta Siemiginowska, 
Manuel Meyer, 
Javier Pascual-Granado,
Rafael Garrido
and the anonymous referee.
Phil Uttley, and 
Simon Vaughan provided 
especially useful comments and suggestions.
I am indebted to the NASA 
Astrophysics Data Analysis program for support
through grant NNX16AL02G,
as well as the NASA 
Applied Information
Systems Research program.



\begin{thebibliography}{99}

\bibitem[Alston, Fabian, Buisson et al.(2019)]{alston_et_al}
Alston, W., Fabian, A., Buisson, D., et al. 2019,
\ifisapj \else
The remarkable X-ray variability of IRAS 13224-3809  I. The variability process, \fi
\mnras, 482, 2088-2106

\bibitem[Alston(2019)]{alston}
Alston, W. 2019,
\ifisapj \else
Non-stationary variability in accreting compact objects, \fi
\mnras, 485, 260-265

\bibitem[Aschwanden(2011)]{aschwanden}
Aschwanden, M. 2011,
Self-Organized Criticality in Astrophysics:
The Statistics of Nonlinear Processes 
in the Universe,
Springer-Verlag, Berlin.


\bibitem[Bar-Lev and Enis(1988)]{barlev_enis}
Bar-Lev, S. and  and Enis, P. 1988,
\ifisapj \else
On the classical choice of variance stabilizing
transformations and an application for a Poisson variate, \fi
Biometrica, 75, 803-804


\bibitem[Bhatta and Dhital(2019)]{bhatta_dhital}
Bhatta, G. and Dhital, N. 2019,
\ifisapj \else
Nature of gamma-ray variability in blazars, \fi
arXiv: 1911.08198

\textbf{\emph{
\bibitem[Box and Cox(1964)]{box_cox}
Box, G. E. P. and Cox, D. R. 1964,
\ifisapj \else
An analysis of transformations, \fi
Journal of the Royal Statistical Society, Series B, 26, 211-252.}}

\bibitem[Brockwell and Davis(1987)]{brockwell_davis}
Brockwell, P. and  and Davis, R. 1987,
Time Series: Theory and Methods,
Springer-Verlag New York, Inc.


\bibitem[Buchler and Kandrup(1997)]{buchler}
Buchler, J.  and Kandrup, H., eds.1997,
Nonlinear Signal and Image Analysis,
Vol. 808,
Annals of the New York Academy of Sciences: 
New York


\bibitem[Crow and Shimizu(1988)]{log_normal_book}
Crow, E. and Shimizu, K. 1988,
Lognormal Distributions: Theory
and Applications,
Marcel Dekker, Inc.: New York


\bibitem[Curtiss(1942)]{curtiss}
Curtiss, J.1942,
\ifisapj \else
On transformations used in the analysis of variance, \fi
Ann. Math. Statist., 14, 107-122


\bibitem[Denis et al.(1994)]{denis}
Denis, W., Olive, J.-F., Roques, J. et al. 1994, 
\ifisapj \else
Noise variability of the hard X-ray transient Nova Persei \fi
ApJS, 92, 459-463.

\bibitem[Dobrotka and Ness(2015)]{dobrotka_ness}
Dobrotka, A. and Ness, J. 2015, 
\ifisapj \else
Differences in the fast optical variability of the dwarf nova V1504 Cyg between quiescence and outbursts detected in Kepler data and simulations of the RMS-flux relations, \fi
\mnras, 451, 2851

\bibitem[Dobrotka, Negoro and Mineshige(2019)]{dobrotka}
Dobrotka, A., 
Negoro, H., and Mineshige, S. 2019,
\ifisapj \else
Similar shot profile morphology of fast variability in
cataclysmic variable, X-ray binary and blazar; the MVLyr case, \fi
arXiv: 1908.11745, 
to appear in A\&A

\bibitem[Eckmann and Ruelle(1992)]{eckmann_ruelle}
Eckmann, J. and Ruelle, D. 1992,
\ifisapj \else
Fundamental limitations for estimating dimensions and Lyapunov exponents in dynamical systems, \fi
Physica D, 56, 185-187


\bibitem[Edelson, Mushotzky, Vaughan et al.(2013)]{edelson}
Edelson, R., Mushotzky, R., Vaughan, S.,
Scargle, J., Gandhi, P., Malkan, M., and  Baumgartner, W. 2013, 
\ifisapj \else
Kepler observations of rapid optical variability 
in the BL Lac object W2R1926+42, \fi
\apj,  766, 16

\bibitem[Efron(1982)]{efron}
Efron, B. 1982,
\ifisapj \else
Transformation Theory: How Normal 
is a Family of Distributions?,
The 1981 Wald Memorial Lectures, \fi
Annals of Statistics, 10, 323-339

\bibitem[Fenimore, Madras, and Nayakshin(1996)]{fenimore}
Fenimore, E., 
Madras,  C.,
and Nayakshin, S. 1996,
\ifisapj \else
Expanding Relativistic Shells and Gamma-Ray Burst
Temporal Structure, \fi
\apj, 473, 998

\bibitem[Giebels and Degrange(2009)]{giebels_degrange}
Giebels, B. and Degrange, B. 2009,
\ifisapj \else
Lognormal variability in BL Lacertae (Research Note) \fi
A\&A, 503, 797-799

\bibitem[Granger and Anderson(1978)]{granger_andersen}
Granger, W. and Anderson, A. 1978,
An Introduction to Bilinear Time Series Models,
Vandenhoeck and Ruprecht: G{\"o}ttingen

\bibitem[Granger and Joyeux(1980)]{granger_joyeux}
Granger, C. and 
Joyeux, R. 1980,
\ifisapj \else
An Introduction to Long Memory Time Series Models and Fractional Differencing, \fi
Journal of Time Series Analysis, 1, 15-30.

\bibitem[Heil, Vaughan and Uttley(2012)]{heil_vaughan_uttley}
Heil, L., Vaughan, S., and Uttley, P. 2012,
\ifisapj \else
The Ubiquity of the RMS-flux relation 
in Black Hole X-ray Binaries, \fi
\mnras, 422, 2620
 
 \bibitem[Hogg and Reynolds(2016)]{hogg_reynolds}
 Hogg, J. and Reynolds, C. 2016,
  \ifisapj \else
 Testing the Propagating Fluctuations Model with a Long, Global Accretion Disk Simulation, \fi
\apj,  826, 40

\bibitem[Kelly et al.(2009)]{kelly_2009} 
Kelly, B., Bechtold, J., \& Siemiginowska, A.\ 2009, 
 \ifisapj \else
Are the Variations in Quasar Optical Flux Driven by Thermal Fluctuations? \fi
\apj, 698, 895

\bibitem[Kelly, Sobolewska, and Siemiginowska(2011)]{kelly_2011}
Kelly, B., Sobolewska, M., 
and Siemiginowska, A. 2011,
 \ifisapj \else
A Stochastic Model for the Luminosity 
Fluctuations of Accreting Black Holes, \fi
\apj,  730, 52

\bibitem[Kelly et al.(2014)]{kelly_2} 
Kelly, B., Becker, A, Sobolewska, M., et al.\ 2014,
 \ifisapj \else
Flexible and Scalable Methods for Quantifying Stochastic 
Variability in the Era of Massive Time-Domain Astronomical Data Sets, \fi
 \apj, 788, 33

\bibitem[Koen(2016)]{koen}
Koen, C. 2016,
 \ifisapj \else
A simple explanation of the linear RMS-mean flux
relation in accreting objects, \fi
A\&A, 593, L17

\bibitem[Kushwaha, Sinha, Misra, Kingh, and
de Gouveia Dal Pino(2017)]{kushwaha}
Kushwaha, P.  
Sinha, A.,  
Misra, R.  
Singh, K.,
de Gouveia Dal Pino, E.  2017,
 \ifisapj \else
Gamma-ray Flux Distribution and Non-linear behavior of Four LAT Bright AGNs, \fi
\apj,  849, 138


\bibitem[Lyubarskii(1997)]{lyubarskii}
Lyubarskii, Y. 1997,
 \ifisapj \else
Flicker noise in accretion disks, \fi
\mnras, 292, 679-685



\bibitem[Meyer, Scargle and Blandford(2019)]{meyer}
Meyer, M., Scargle, J. and Blandford, R. 2019,
 \ifisapj \else
Characterizing the Gamma-Ray Variability 
of the Brightest Flat Spectrum Radio Quasars Observed with the Fermi LAT, \fi
\apj, 877, 39.

\bibitem[Mineshige, Ouchi and Nishimori(1994)]{mineshige}
Mineshige, S., Ouchi, N., 
and Nishimori, H. 1994,
 \ifisapj \else
On the Generation of 1/f Fluctuations
in X-Rays from Black-Hole Objects, \fi
Publ. Astron. Soc. Japan, 46, 97-105

 

\bibitem[Nemiroff, Norris, Kouveliotou, et al.(1994)]{nemiroff}
Nemiroff, R.,
Norris, J.,
Kouveliotou, C.,
Fishman, G.,
 Meegan, C.,
 and 
 Paciesas, W. 1994,
 \ifisapj \else
 Gamma-Ray Bursts are Time-Asymmetric \fi
\apj 423, 432


\bibitem[Osborne and Provenzale(1989)]{osborne_provenzale}
Osborne, A.  and Provenzale, A. 1989,
\ifisapj \else
Finite correlaiton dimension 
for stochastic systems with 
power-law spectra, \fi
Physica D 47, 361-372


\bibitem[Parzen(1962)]{parzen}
Parzen,E., 
\ifisapj \else
Spectral analysis of asymptotically 
stationary time series, \fi
Bull. Inst. Internat. Statist., 39, 87-2013, 1962


\bibitem[Pascual-Granado, 
Garrido and Su\'arez(2013)]{javi_rafa_1}
Pascual-Granado, J.,
Garrido, R.,
Su\'arez, J. 2013,
\ifisapj \else
On the necessity of a new interpretation of the stellar light curves, \fi
Proceedings of the IAU, Vol. 9, S301, 2013, pp 85-88
arXiv:1311.3553	

\bibitem[Pascual-Granado, 
Garrido and Su\'arez(2015)]{javi_rafa_2}
Pascual-Granado, J.,
Garrido, R.,
Su\'arez, J. 2015,
\ifisapj \else
Inconsistencies in the application of harmonic analysis to pulsating stars,
Astronomy and Astrophysics, \fi
581, A89,
arXiv:1507.07877


\bibitem[Phillipson, Boyd and Smale(2018)]{phillipson}
Phillipson,  R., Boyd, P. and Smale, A. 2018,
\ifisapj \else
The Chaotic Long-term X-ray Variability of 4U 1705-44, \fi
\mnras, 477, 5220-5237


\bibitem[Press(1978)]{press}
Press, W.1978, 
\ifisapj \else
Flicker Noise in Astronomy and Elsewhere, \fi
Comments Astrophys, 7, 103-119.

\bibitem[Priestley(1988)]{priestley}
Priestley, M. 1988,
Non-linear and Non-Stationary Time Series Analysis,
Academic Press: New York

\bibitem[Ruelle(1990)]{ruelle}
Ruelle, D. 1990,
\ifisapj \else
Deterministic chaos: the science and the fiction,
The Claude Bernard Lecture, \fi
1989,
Proc. R. Soc. Lond. A, 427, 241-248


\bibitem[Scargle(1981a)]{scargle_findley}
Scargle, J. 1981a,
\ifisapj \else
Phase-sensitive deconvolution to model random
processes, with special reference to 
astronomical data, \fi
1981, In  Applied Time 
Series Analysis II, D. Findley ed. New York: Academic Press.


\bibitem[Scargle(1981b)]{scargle_i}
Scargle, J. 1981b,
\ifisapj \else
Studies in Astronomical Time Series Analysis: I. Modeling Random Processes in the Time Domain, \fi
\apjs,  45, 1


\bibitem[Scargle, Donoho, Crutchfield et al.(1992)]{scargle_dhr}
Scargle, J.,
Donoho, D.,
Crutchfield, J.,
Steiman-Cameron,  T., 
Imamura, J., 
 and K. Young, K. 1993,
 \ifisapj \else
The Quasi-Periodic Oscillations and Low-Frequency Noise of Scorpius X-1 as
Transient Chaos; A Dripping Handrail?, \fi
Astrophysical Journal Letters,
411, L91-L94.


\bibitem[Scaringi et al.(2012)]{scaringi_1}
Scaringi, S., K\"ording, Uttley, P., 
Knigge, C., Groot, P. and Still, M. 2012,
\ifisapj \else
The Universal Nature of Accretion-induced 
Variability:
the RMS-Flux Relation in an Accreting White Dwarf, \fi
\mnras, 421, 2854

\bibitem[Shah et al.(2018)]{shah}
Shah, Z., 
Mankuzhiyil,  N.,
Sinha, A.,
Misra, R.,
Sahayanathan, S., and 
Iqbal, N. 2018,
\ifisapj \else
Log-normal flux distribution of bright Fermi blazars, \fi
Research in Astronomy and Astrophysics, 18, 141


\bibitem[Sinha, Khatoon, Mistra at al.(2018)]{sinha}
Sinha, A., Khatoon, R., Misra, R., Sahayanathan, S. 
Mandal, S., Gogoi, R., Bhatt, N. 2018, 
\ifisapj \else
The flux distribution of individual blazars as a key to
understand the dynamics of particle acceleration, \fi
\mnras \ Letters, 480, L 116-120

\bibitem[Smith et al.(2018)]{krista}
Smith, K., Mushotzky, R., Boyd, P.,
Malkan, M., Howell, S. and 
Gelino, D. 2018,
\ifisapj \else
The Kepler Light Curves of AGN: A Detailed Analysis, \fi
\apj,  857, 141



\bibitem[Sprott(2003)]{sprott}
Sprott, J. 2003,
Chaos and Time-Series Aanalysis,
Oxford University Press, Oxford

\bibitem[Takata, Mukuto and Mizumoto(2018)]
{takata}
Takata,T., Mukuta, Y, and  Mizumoto, Y. 2018,
\ifisapj \else
Modeling the Variability of Active Galactic Nuclei by an Infinite Mixture of Ornstein-Uhlenbeck (OU) Processes, \fi
\apj,  869, 178
[Erratum: 2018, ApJ, 869, 178]
 
\bibitem[Theiler, Lindday and Rubin(1993)]{theiler}
Theiler, J., Lindday, P. and Rubin, D. 1993,
\ifisapj \else
Detecting Nonlinearity in Data
with Long Coherence Times, \fi
in Predicting the Future
and Understanding the Pose, Eds. Weigent, A. and Gerschenfeld, N.,
SFI Studies in the Sciences of Complexity,
Proc. Vol. XVII, Addison-Wesley,


\bibitem[Thorne and Blandford(2017)]{thorne_blandford}
Thorne, K. and Blandford, R. 2017,
Modern Classical Physics,
Princeton University Press

\bibitem[Tong(1990)]{tong}
Tong, H. 1990,
Non-linear Time Series: 
A Dynamical System Approach,
Clarendon Press: Oxford


\bibitem[Uttley and McHardy(2001)]{uttley_mchardy}
Uttley, P., and
McHardy, I.,
2001,
\ifisapj \else
The flux-dependent amplitude of broadband
noise variability in X-ray binaries and active galaxies, \fi
\mnras \ 323, L26.

\bibitem[Uttley et al.(2005)]{uttley_mchardy_vaughan}
Uttley, P., McHardy, I., and Vaughan, S. 2005,
\ifisapj \else
Non-linear X-ray variability 
in X-ray binaries and active galaxies, \fi
\mnras, 359, 345-362


\bibitem[Uttley, McHardy and Vaughan(2017)]{uttley_4}
Uttley, P., 
McHardy, I.,
and Vaughan, S.
2017, 
\ifisapj \else
The RMS-flux relation in accreting objects:
not a simple "volume control," \fi
A\&A, 601, 1

\bibitem[Vaughan and Uttley(2008)]{vaughan_uttley}
Vaughan, S. and Uttley, P. 2008,
\ifisapj \else
Studying accreting black holes and neutron stars 
with time series: beyond the power specrum, 
``Noise and Fluctuations," \fi
Proc. SPIE, 6603

\bibitem[Vaughan(2013)]{vaughan}
Vaughan, S. 2013,
\ifisapj \else
Random time series in Astronomy, \fi
Phil. Trans. R. Soc. A, 1-28


\bibitem[Wold(1938)]{wold}
Wold, Herman O. A. 1938,
A Study in the Analysis of Stationary Time Series,
Almqvist \& Wiksells: Uppsala

\bibitem[Young and Scargle(1996)]{young}
Young, K. and Scargle, J. 1996,
\ifisapj \else
The Dripping Handrail Model: Transient Chaos in Accretion Systems \fi
\apj, 468, 617


\bibitem[Yule(1926)]{yule}
Yule, G. 1926,
\ifisapj \else
Why do we Sometimes get Nonsense Correlations
between Time Series? -- A Study in
Sampling and the Nature of Time-Series, \fi
Journal of the Royal Statistical Society,
89, 1-63.


\end{thebibliography}
\end{document}